\documentclass[twocolumn,letter]{jpsj3}
\usepackage{txfonts}
\usepackage{graphicx}
\usepackage{dcolumn}
\usepackage{bm}
\usepackage[usenames,dvipsnames]{xcolor}
\usepackage{ulem}
\usepackage{amsmath}
\title{
Asymmetric Magnon Excitation by Spontaneous Toroidal Ordering
}

\author{Satoru Hayami$^{1,2}$\thanks{Present institution, Department of Physics, Hokkaido University, Sapporo 060-0810, Japan}, Hiroaki Kusunose$^3$, and Yukitoshi Motome$^2$}
\inst{
$^1$Theoretical Division, Los Alamos National Laboratory, Los Alamos, New Mexico 87545, USA \\
$^2$Department of Applied Physics, University of Tokyo, Tokyo 113-8656, Japan \\
$^3$Department of Physics, Meiji University, Kanagawa 214-8571, Japan 
 }

\abst{
Effects of spontaneous toroidal ordering on magnetic excitation are theoretically investigated for a localized spin model that includes a staggered Dzyaloshinsky-Moriya interaction and anisotropic exchange interactions, which arise from the antisymmetric spin-orbit coupling and the multi-orbital correlation effect. 
We show that the model exhibits a N\'eel-type antiferromagnetic order, which simultaneously accompanies a ferroic toroidal order. 
We find that the occurrence of toroidal order modulates the magnon dispersion in an asymmetric way with respect to the wave number: 
a toroidal dipole order on the zigzag chain leads to a band-bottom shift, while a toroidal octupole order on the honeycomb lattice gives rise to a valley splitting. 
These asymmetric magnon excitations could be a source of unusual magnetic responses, such as nonreciprocal magnon transport. 
A variety of modulations are discussed while changing the lattice and magnetic symmetries. 
Implications of candidate materials for asymmetric magnon excitations are presented. 
}

\begin{document}
\maketitle

Effects of the spin-orbit coupling (SOC) on magnetic materials have attracted a great interest for the rich physics emergent from entangled spin and orbital degrees of freedom. 
Magnetoelectric effects caused by coexistence of ferromagnetic (FM) and ferroelectric orders are typical examples~\cite{curie1894symetrie,Fiebig0022-3727-38-8-R01,KhomskiiPhysics.2.20}. 
Another examples are magnetotransport properties, such as the topological Hall effect in magnetic skyrmions~\cite{nagaosa2013topological}. 
In these phenomena, the lack of spatial inversion symmetry plays an important role: 
the SOC acquires an antisymmetric component with respect to the wave number, which is called the antisymmetric spin-orbit coupling (ASOC). 

The ASOC can exist even in centrosymmetric systems in a hidden form. 
Suppose the inversion center is located at an off-site position, and the inversion symmetry is broken at each lattice site. 
Then, the local asymmetry may induce an ASOC in a sublattice-dependent form, whose net component vanishes~\cite{zhang2014hidden}. 
For instance, this class of centrosymmetric systems are found in zigzag chain, honeycomb, and diamond lattices.

In such situations, interesting physics arises once a spontaneous breaking of sublattice symmetry takes place by electron correlations: 
a net component of ASOC is induced by the additional inversion symmetry breaking~\cite{Yanase:JPSJ.83.014703,Hayami_PhysRevB.90.024432,Hayami_PhysRevB.90.081115,Hayami_doi:10.7566/JPSJ.84.064717,Fu_PhysRevLett.115.026401}. 
This opens the possibility of controlling the ASOC through the electronic degrees of freedom. 
Furthermore, such an emergent ASOC has a close relationship to odd-parity multipoles. 
For instance, in the one-dimensional (1D) zigzag-chain system where the ASOC is present in the staggered form, a N\'eel-type antiferromagnetic (AFM) order induces an additional net ASOC. 
This accompanies a ferroic order of odd-parity multipoles, such as magnetic toroidal and quadrupole, which break both spatial and time-reversal symmetries~\cite{Yanase:JPSJ.83.014703,Hayami_doi:10.7566/JPSJ.84.064717}. 
Interestingly, such a toroidal order modifies the electronic band structure in an asymmetric way in the momentum space~\cite{volkov1981macroscopic}, resulting in unusual off-diagonal responses, such as the magneto-current effect~\cite{Yanase:JPSJ.83.014703,Hayami_PhysRevB.90.024432}. 

Spontaneous toroidal ordering in the presence of ASOC is also expected to 
affect dynamical properties, such as a nonreciprocal directional dichroism, through the collective excitations. 
Moreover, as the odd-parity multipoles are associated with conventional orders, such as the N\'eel AFM, they are realized without magnetic frustration that usually suppresses the relevant energy scale. 
Hence, they have a potential for next-generation electronic devices working at higher  temperatures. 
Nevertheless, the microscopic theory has not been fully elaborated thus far.

In the present paper, we investigate the magnon excitations in N\'eel-type AFM ordered states in the centrosymmetric systems with local asymmetry. 
Considering the Heisenberg model with a staggered Dzyaloshinsky-Moriya (DM) interaction~\cite{dzyaloshinsky1958thermodynamic,moriya1960anisotropic,Starykh_PhysRevB.82.014421} and anisotropic exchange interactions~\cite{kaplan1983single,Shekhtman_PhysRevB.47.174}, which originate from the ASOC and multi-orbital correlation effect, we show that the AFM orders with ferroic toroidal ordering can be realized. 
We find that the spontaneous odd-parity multipole orders result in peculiar asymmetric deformations of the magnon dispersion, such as a shift of the band bottom and a valley splitting. Such deformations lead to unconventional off-diagonal responses even in the absence of external fields. 
Extending the analysis to other cases including FM and locally symmetric cases, we systematically elaborate how the magnon excitation is modulated by the lattice and magnetic symmetries. 

\begin{figure}[t]
\begin{center}
\includegraphics[width=1.0 \hsize]{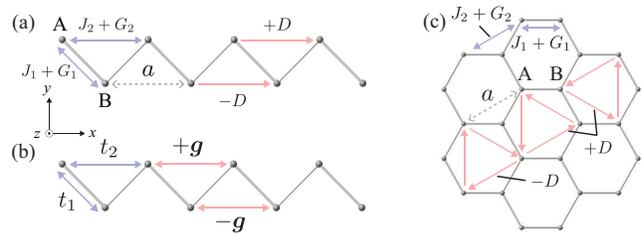} 
\caption{
\label{Fig:Zigzag_ponti_souzu}
(Color online)
Schematic pictures of the zigzag chain along the $x$ direction for (a) the localized spin model in Eq.~(\ref{eq:HamSpin}) and (b) the itinerant electron model in Eq.~(\ref{eq:Ham_Hubbard}).  
(c) Schematic picture of the honeycomb lattice for the localized spin model in the $xy$ plane.  
In (a) and (c), A and B represent two sublattices. 
}
\end{center}
\end{figure}

Let us begin with the 1D zigzag chain, the simplest realization of the centrosymmetric lattices with local asymmtry. 
We consider the Heisenberg model supplemented by anisotropic exchange interactions and a sublattice-dependent DM interaction, whose Hamiltonian is given by 
\begin{align}
\label{eq:HamSpin}
\mathcal{H}&= \sum_{i,j } J_{ij} \bm{S}_i \cdot \bm{S}_j
+ \sum_{ i,j } G_{ij} \left( S_i^{z} S_j^{z} -S_i^{x} S_j^{x} -S_i^{y} S_j^{y}  \right)
\nonumber \\
&+ D\sum_{p} \left[
\left(\bm{S}_{{\rm A}p} \times \bm{S}_{{\rm A}p+1} \right)-\left(\bm{S}_{{\rm B}p} \times \bm{S}_{{\rm B}p+1} \right)\right]\cdot\bm{z},
\end{align}
where $\bm{S}_i$ is the $S=1/2$ operator at site $i=(l,p)$ ($l$ and $p$ denote the indices for the sublattice A and B, and unit cell, respectively). 
The first term represents the Heisenberg-type exchange coupling for nearest-neighbor (NN) and next-nearest-neighbor (NNN) spins, $J_1$ and $J_2$, respectively. 
The second term in Eq.~(\ref{eq:HamSpin}) is the anisotropic exchange interactions~\cite{kaplan1983single,Shekhtman_PhysRevB.47.174} for NN and NNN spins, $G_1$ and $G_2$, respectively. 
The former $G_1$ originates from the multi-orbital correlation effect, while the latter $G_2$ from the ASOC, as discussed below. 
The third term in Eq.~(\ref{eq:HamSpin}) is the antisymmetric exchange interaction between NNN spins (DM interaction)~\cite{dzyaloshinsky1958thermodynamic,moriya1960anisotropic,Starykh_PhysRevB.82.014421} [see Fig.~\ref{Fig:Zigzag_ponti_souzu}(a)], whose origin is also the ASOC as shown below. 
Note that there is no DM interaction for NN spins because of the inversion center at the bond median. 

In the second and third terms in Eq.~(\ref{eq:HamSpin}), the sublattice-dependent ASOC inherited from the local asymmetry of the zigzag lattice plays an important role. 
This is understood, for instance, by considering the strong-coupling limit of an effective single-band model~\cite{Yanase:JPSJ.83.014703}, whose Hamiltonian is given by 
\begin{align}
\label{eq:Ham_Hubbard}
\mathcal{H}&=
\sum_{i,j,\sigma} t_{ij} (c^{\dagger}_{ i \sigma} c^{}_{j \sigma} + {\rm H.c.})+U \sum_i n_{i\uparrow} n_{i \downarrow} \nonumber \\
 &+  \sum_{k, \sigma , \sigma'} \bm{g} (k) \cdot \bm{\sigma}_{\sigma \sigma'} \left( c^{\dagger}_{{\rm A} k  \sigma} c^{}_{{\rm A} k \sigma'} - c^{\dagger}_{{\rm B} k  \sigma} c^{}_{{\rm B} k \sigma'} \right). 
\end{align}
The first and second terms comprise the Hubbard model in a standard notation; $t_1$ and $t_2$ are NN and NNN hoppings, respectively. 
The third term represents the sublattice-dependent ASOC in the form of the spin-dependent electron hopping, which originates from the atomic SOC, local asymmetry, and multi-orbital hybridization effects between orbitals with different parity~\cite{Yanase:JPSJ.83.014703,Hayami_PhysRevB.90.024432,sugita2015antisymmetric}. 
$\bm{g}$ is the asymmetric vector with respect to $k$: $\bm{g} (k)= 2\alpha\sin(ka)\bm{z}$ [$\alpha$ is the amplitude, $a$ is the lattice constant ($a=1$), and $\bm{z}$ is the unit vector along the $z$ axis; see Figs.~\ref{Fig:Zigzag_ponti_souzu}(a) and \ref{Fig:Zigzag_ponti_souzu}(b)], and $\bm{\sigma}$ is the Pauli matrix. 
The sign factor comes from the zigzag structure, as an electron at the A and B sublattice sites feels an opposite potential gradient. 
By considering the strong-coupling limit of the model in Eq.~(\ref{eq:Ham_Hubbard}) at half-filling, we obtain the coupling constants in the model in Eq.~(\ref{eq:HamSpin}) by the second-order perturbation in terms of $t_m/U$ ($m=1,2$) and $\alpha/U$ as $J_m=4 t_m^2/U$, $G_2=4\alpha^2/U$, and $D=2\sqrt{J_2 G_2}$. 
On the other hand, in general, the multi-orbital correlation effect, which is not taken into account in Eq.~(\ref{eq:Ham_Hubbard}), favors the N\'eel-type AFM ordering along the $z$ direction in the presence of the atomic SOC and the multi-orbital hybridization, as demonstrated in a $d$-$p$ model~\cite{comment_Sugita} and an extended Kondo lattice model~\cite{Hayami_doi:10.7566/JPSJ.84.064717}. 
Such a multi-orbital correlation effect is incorporated in the $G_1$ term in Eq.~(\ref{eq:HamSpin}). 
Hereafter, we consider $J_1>0$ and $J_2 \geq 0$ with taking $J_1$ as an energy unit, and $G_1\geq0$, $G_2\geq0$, and $D=2\sqrt{J_2 G_2}$. 

\begin{figure}[t]
\begin{center}
\includegraphics[width=1.0 \hsize]{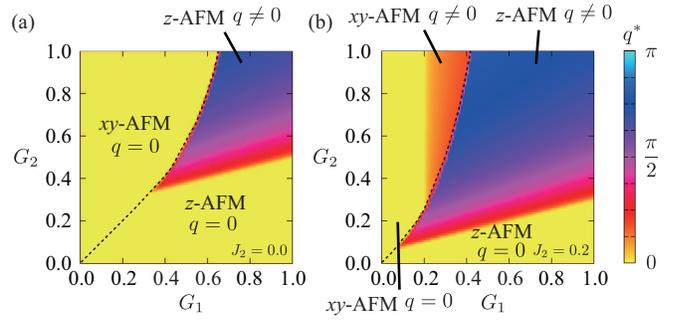} 
\caption{
\label{Fig:Zigzag_phase_diagram}
(Color online)
The ground-state phase diagrams for the model in Eq.~(\ref{eq:HamSpin}) at (a) $J_2=0$ and (b) $J_2=0.2$. 
The contour displays the ordering wave number $q^*$. 
$xy$($z$)-AFM represents the AFM state with the spin polarization along the $xy$($z$) direction.
}
\end{center}
\end{figure}

First, we examine the ground state of the model in Eq.~(\ref{eq:HamSpin}). 
To stabilize the long-range orders, we here consider a three dimensional system composed of ferromagnetically weakly- coupled 1D zigzag chain.
We obtain the ground-state phase diagram by the Luttinger-Tisza method by treating the spins as classical vectors~\cite{Luttinger_PhysRev.70.954}. 
The results are plotted in Figs.~\ref{Fig:Zigzag_phase_diagram}(a) and \ref{Fig:Zigzag_phase_diagram}(b) for $J_2=0$ and $J_2=0.2$, respectively.
The contour displays the optimal wave number $q^*$. 
When $J_2=G_1=G_2=0$ [the origin in Fig.~\ref{Fig:Zigzag_phase_diagram}(a)], the stable spin configuration is a $q=0$ collinear AFM without any spin anisotropy. 
With increasing $G_1$ ($G_2$), the NN (NNN) anisotropic interaction stabilizes the AFM ordering along the $z$ direction ($z$-AFM) [in the $xy$ plane ($xy$-AFM)] without any spin canting. 
Note that the system has the spin-rotational symmetry around the
$z$ axis. 
With further increasing $G_1$ and $G_2$, incommensurate magnetic orders with longer period $q \neq 0$ appear. 
As shown in Fig.~\ref{Fig:Zigzag_phase_diagram}(b), the $q \neq 0$ region becomes larger for larger $J_2$, as there is frustration between $J_1$ and $J_2$.  

The results in Figs.~\ref{Fig:Zigzag_phase_diagram}(a) and \ref{Fig:Zigzag_phase_diagram}(b) show that the model exhibits the collinear $z$-AFM order for sufficiently large $G_1/G_2$. 
As described above, the $z$-AFM order on the zigzag chain is of particular interest as it accompanies the odd-parity multipoles, such as magnetic toroidal dipoles and magnetic quadrupoles, through the spin-orbit coupling effect when $D \neq 0$. 
The magnetic toroidal dipoles we are considering in this study are defined by $\bm{t}=\sum_{i}(\bm{r}_i \times \bm{S}_i)$, where $\bm{r}_i$ is the position vector from the inversion center to the lattice site $i$~\cite{Spaldin:0953-8984-20-43-434203}.
In such a situation, the $z$-AFM order possess the toroidal dipoles in the $x$ direction because $\bm{S}_i \parallel \bm{z}$ and $\bm{r}_i \parallel \bm{y}$ by taking the bond median as the origin.

Next, we examine magnetic excitations in the collinear $z$-AFM state by using the spin-wave theory. 
We adopt the standard Holstein-Primakoff transformation, which is represented by $S_{{\rm A}q}^{+}= \sqrt{2S}a_{q}$ ($S_{{\rm B}q}^{+}=\sqrt{2S}b_{q}^{\dagger}$), $S_{{\rm A}q}^{-} =\sqrt{2S} a_{q}^{\dagger}$ ($S_{{\rm B}q}^{-} =\sqrt{2S} b_{q}$), and $S_{{\rm A}q}^z =S-a_{q}^{\dagger}a_{q}$ ($S_{{\rm B}q}^z =-S+b^{\dagger}_{q}b_{q}$) for the A (B) sublattice. 
Here, $a$ and $b$ are the boson operators for the A and B sublattices, respectively, and $q$ is the wave number along the chain direction. 
We adopt the linear spin-wave approximation, in which the magnon-magnon interactions are ignored.
Then, the spin-wave Hamiltonian in the $q$ space is obtained as 
\begin{align}
\label{eq:Ham_spnwave}
\mathcal{H} = E_0 &
+ \sum_q  (A^{{\rm S}}_q + A^{{\rm AS}}_q)  (a^{\dagger}_q a_q^{} + b^{\dagger}_q b_q^{}) \nonumber  \\
&+ \sum_q B_q (a_q b_{-q} + a^{\dagger}_q b_{-q}^{\dagger}),
\end{align}
where $A_q^{{\rm S}}  = 2S \left[ (J_1 + G_1)  + J_2  (\cos q -1)  - G_2  (\cos q +1 ) \right]$, $A_q^{{\rm AS}} =2 SD\sin q$, $B_q = 2 S (J_1- G_1) \cos (q/2)$, and $E_0=-NS^2 (J_1 -J_2 + G_1 -G_2)$. 
By using the Bogoliubov transformations $\alpha_q = a_q \cosh \theta_q + b^{\dagger}_{-q} \sinh \theta_q$ and $\beta_{-q}^{\dagger} = a_q \sinh \theta_q + b^{\dagger}_{-q} \cosh \theta_q$ [$\theta_q = (1/2)\tanh^{-1} (B_q/A_q^{{\rm S}})$], we can diagonalize Eq.~(\ref{eq:Ham_spnwave}) into the form of 
$\mathcal{H} = \sum_q \omega_q (\alpha_q^{\dagger} \alpha_q^{} + \beta_q^{\dagger} \beta_q^{})  + {\rm const.}$, where the magnon dispersion relation $\omega_q$ is obtained as 
\begin{align}
\label{eq:magnon_dis}
\omega_q= \sqrt{(A^{{\rm S}}_q)^2 - B_q^2} + A_q^{{\rm AS}}. 
\end{align}

\begin{figure}[t]
\begin{center}
\includegraphics[width=1.0 \hsize]{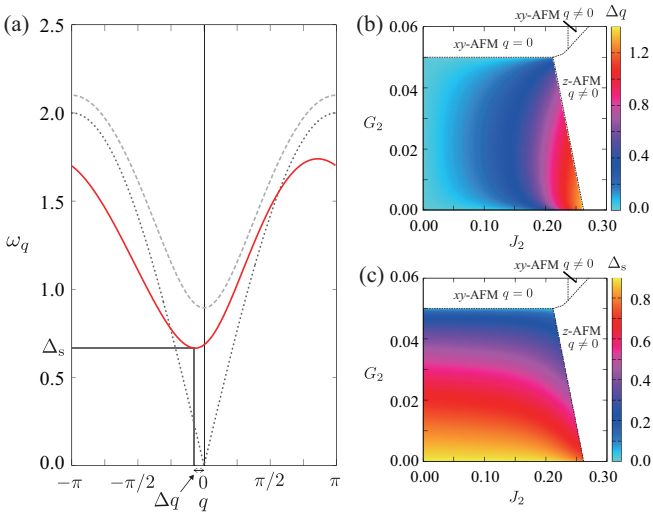} 
\caption{
\label{Fig:Zigzag_spin_wave}
(Color online)
(a) Magnon dispersion in Eq.~(\ref{eq:magnon_dis}) in the $z$-AFM state on the zigzag chain. 
The result is calculated at $(J_2, G_1, G_2)=(0.1, 0.05, 0.02)$. 
For comparison, the results for $(J_2, G_1, G_2)=(0, 0, 0)$ and $(J_2, G_1, G_2)=(0, 0.05, 0)$ are shown by the dotted and dashed curves, respectively. 
All the dispersions are doubly degenerate. 
$J_2$-$G_2$ plots of (b) $\Delta q$ and (c) $\Delta_{\rm s}$ at $G_1=0.05$. 
The blank regions represent the areas where the $xy$-AFM and other incommensurate orders appear. 
}
\end{center}
\end{figure}

Figure~\ref{Fig:Zigzag_spin_wave}(a) shows the magnon dispersion in the $z$-AFM state at $J_2=0.1$, $G_1=0.05$, and $G_2=0.02$. 
The magnon excitation spectrum has a peculiar form: 
in addition to the gap opening, the dispersion undergoes an asymmetric deformation with respect to $q$, leading to a shift of the band bottom from $q=0$. 
This is similar to the electronic band structure in the presence of the toroidal ordering~\cite{Yanase:JPSJ.83.014703,Hayami_PhysRevB.90.024432}. 
For comparison, we also show the results at $J_2=G_1=G_2=0$ (dotted lines) and $J_2=G_2=0$ (dashed lines). 
The results indicate that $G_1$ opens the excitation gap and $D = 2\sqrt{J_2 G_2}$ brings about the asymmetric deformation. 

We analyze the effect of $G_1$, $G_2$, and $J_2$ more carefully. 
Figures~\ref{Fig:Zigzag_spin_wave}(b) and \ref{Fig:Zigzag_spin_wave}(c) represent the band-bottom shift $\Delta q$ and the spin gap $\Delta_{\rm s}$ on the $J_2$-$G_2$ planes at $G_1=0.05$. 
The results show that $J_2$ and $G_2$ significantly affect the asymmetric deformation of the magnon dispersion, as shown in Fig.~\ref{Fig:Zigzag_spin_wave}(b). 
This is because the deformation is caused by $A_q^{\rm AS} \propto D \sin q$ originating from the sublattice-dependent DM interaction in the third term in Eq.~(\ref{eq:HamSpin}). 
On the other hand, the spin gap depends on both $G_1$ and $G_2$, while it does not show strong $J_2$ dependence, except in the region close to an incommensurate phase, as shown in Fig.~\ref{Fig:Zigzag_spin_wave}(c).

\begin{figure}[t]
\begin{center}
\includegraphics[width=1.0 \hsize]{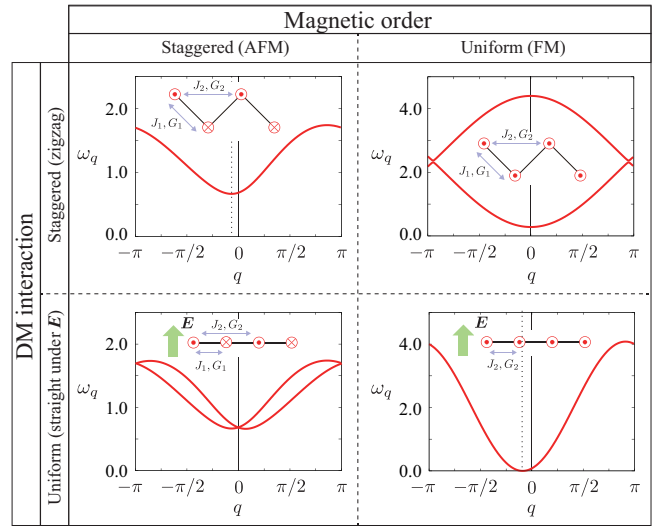} 
\caption{
\label{Fig:Table_magnon_dispersion}
(Color online) 
Summary of the characteristic of the magnon dispersions on the 1D systems. 
The row represents staggered- or uniform-type DM interaction, while the column represents staggered- or uniform-type magnetic ordering polarized in the $z$ direction. 
The dispersions are calculated for $(J_1, J_2, G_1, G_2)= (1, 0.1, 0,05, 0.02)$ (upper left), $(1, 0.1, 0.05, 0.02)$ (lower left), $(-1, -0.1, -0.05, -0.02)$ (upper right), and $(0, -1, 0, -0.02)$ (lower right) in the model in Eq.~(\ref{eq:HamSpin})~\cite{comment_exchange_int}. 
While the lower left is obtained by assuming the same sign for $D$ (the uniform DM) in Eq.~(\ref{eq:HamSpin}), the upper right is calculated in the FM ordered state.
A schematic picture for each magnetic state is shown in each inset. 
The vertical dotted lines represent the shifted band bottoms. 
}
\end{center}
\end{figure}

For further elucidating the relation between the magnon dispersion and the symmetry of the system, 
we extend our analysis to other cases. 
Namely, we consider both AFM and FM cases while changing the parameters in Eq.~(\ref{eq:HamSpin}) beyond the range expected from the perturbation theory. 
We also consider a straight 1D chain in an applied electric field, which has a uniform DM interaction. 

Figure~\ref{Fig:Table_magnon_dispersion} summarizes the characteristic of the magnon dispersions in the matrix form for the staggered/uniform DM interactions (row) and the AFM/FM orders (column). 
The above result for the spontaneous $z$-AFM ordering on the zigzag chain is shown in the upper-left panel. 
A similar asymmetric dispersion also appears for the $z$-FM ordering under the uniform DM interaction (lower-right)~\cite{Melcher_PhysRevLett.30.349}. 
Note that the latter is in the same category as the toroidal magnon discussed for the helical magnetic structure in an applied magnetic field~\cite{MiyaharaJPSJ.81.023712,Miyahara_PhysRevB.89.195145}. 
These asymmetric magnon dispersions lead to peculiar magnetooptical phenomena, such as the nonreciprocal directional dichroism~\cite{Iguchi_PhysRevB.92.184419,seki2015nonreciprocal}. 

On the other hand, when the $z$-AFM order occurs on the straight chain with the uniform DM, the magnon dispersion undergoes a different modulation (lower-left), which is similar to a ``Rashba-type" splitting in the electronic band structure under a uniform ASOC. 
Meanwhile, in the case of the $z$-FM order on the zigzag chain, there is a symmetric modulation, reflecting the spatial inversion symmetry (upper-right).

\begin{figure}[t]
\begin{center}
\includegraphics[width=1.0 \hsize]{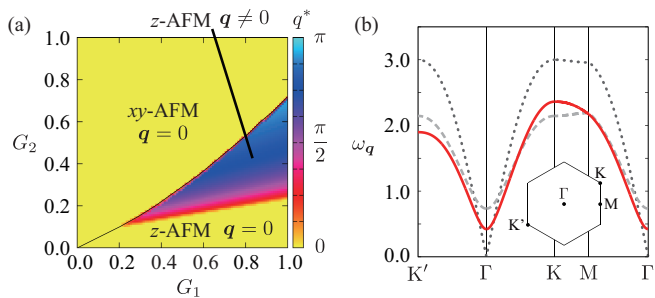} 
\caption{
\label{Fig:Honeycomb}
(Color online) 
(a) The ground-state phase diagram on the honeycomb lattice at $J_2=0.1$. 
The contour displays the ordering wave number $q^* = \sqrt{q_x^2+q_y^2}$. 
(b) Magnon dispersion under the $z$-AFM ordering on the honeycomb lattice. 
The result is calculated at $(J_2, G_1, G_2)=(0.1, 0.015, 0.005)$. 
The dotted and dashed curves represent the magnon dispersions at $(J_2, G_1, G_2)=(0, 0, 0)$ and $(J_2, G_1, G_2)=(0.1, 0.015, 0)$, respectively. 
The inset of (b) shows the Brillouin zone. 
}
\end{center}
\end{figure}

Finally, let us extend the analysis to the two-dimensional case. 
We study the magnon dispersion in the collinear $z$-AFM state on the honeycomb lattice. 
Similar to the 1D zigzag case, the spin model includes the anisotropic and antisymmetric interactions, as shown in Fig.~\ref{Fig:Zigzag_ponti_souzu}(c). 
$D$ is positive (negative) in the counterclockwise (clockwise) direction in the hexagonal plaquettes.
In the case of the honeycomb lattice, a collinear $z$-AFM order accompanies ferroic toroidal octupoles instead of dipoles, reflecting the presence of threefold rotational symmetry of the lattice, i.e., the net toroidal-dipole component is zero. 
We find that, by the Luttinger-Tisza analysis, such a $z$-AFM order with toroidal octupoles becomes stable in the ground state for large $G_1/G_2$, although the region becomes narrower than that for the zigzag chain, as shown in Fig.~\ref{Fig:Honeycomb}(a). 

In the collinear $z$-AFM ordered state, the magnon dispersion does not show any shift of the band bottom, as the lowest contribution to the ASOC is in the third order in $(q_{x},q_{y})$ due to the threefold rotational symmetry. 
Instead, the asymmetry appears near the Brillouin zone boundary around the K and K' points, as shown in Fig.~\ref{Fig:Honeycomb}(b). 
This is a valley splitting, similar to the electronic band structure discussed in noncentrosymmetric compounds, such as monolayer transition metal dichalcogenides~\cite{xiao2012coupled}. 
Note that the band-bottom shift is obtained also on the honeycomb lattice once a uniaxial pressure is applied, since it breaks the threefold rotational symmetry and induces a $q$-linear contribution to the ASOC as in the zigzag chain case. 

To summarize, we have clarified that spontaneous ordering of toroidal multipoles modulates the magnon excitations in an asymmetric way in the momentum space. 
The observation of asymmetric magnon spectra will provide an experimental probe for toroidal multipoles. 
There are many experimental candidates. 
One is the AFM zigzag compound $\alpha$-Cu$_2$V$_2$O$_7$~\cite{Gitgeatpong_PhysRevB.92.024423}. 
Since the compound may possess both the staggered and uniform DM interactions due to the peculiar magnetic and lattice structures, both a band-bottom shift and ``Rashba-type" splitting are expected to be observed. 
Another candidates are transition metal tricalcogenides, such as MnPS$_3$ and MnPSe$_3$~\cite{Ressouche_PhysRevB.82.100408,li2013coupling,Sivadas_PhysRevB.91.235425}: 
the $z$-AFM order on the honeycomb lattice will lead to a valley splitting in the magnon spectrum. 
A distorted honeycomb compound $\beta$-YbAlB$_4$ is also a candidate, as it shows the AFM ordering under a uniaxial pressure~\cite{tomita2015high}; a band-bottom shift is expected in addition to a valley-splitting. 
We note that the diamond-lattice systems will also be in the present scope, as they possess the similar physics related with the local asymmetry. 

Similar asymmetric excitations will be obtained also in itinerant electron systems, such as the Hubbard and Kondo lattice models, when the sublattice-dependent ASOC is present. Reflecting the itinerant electron degree of freedom, asymmetric Stoner-type excitations are expected in addition to asymmetric magnon excitations. Such an analysis is left for future study. 

\begin{acknowledgments}
The authors thank T. Arima for enlightening discussions. 
They also thank K. Matan, Y. Onose, and T. J. Sato for fruitful discussions on the experimental situations. 
Work at LANL was performed under the auspices of the U.S. DOE contract No. DE-AC52- 06NA25396 through the LDRD program.
This work was supported by Grants-in-Aid for Scientific Research (No.~24340076 and 15K05176), Grant-in-Aid for Scientific Research on Innovative Areas (No.~15H05885), the Strategic Programs for Innovative Research (SPIRE), MEXT, and the Computational Materials Science Initiative (CMSI), Japan. 
\end{acknowledgments}

\bibliographystyle{JPSJ}
\bibliography{ref}

\begin{thebibliography}{10}

\bibitem{curie1894symetrie}
P.~Curie, J. Phys. Theor. Appl. {\bfseries 3},  393 (1894).

\bibitem{Fiebig0022-3727-38-8-R01}
M.~Fiebig, J. Phys. D: Appl. Phys. {\bfseries 38},  R123 (2005).

\bibitem{KhomskiiPhysics.2.20}
D.~Khomskii, Physics {\bfseries 2},  20 (2009).

\bibitem{nagaosa2013topological}
N.~Nagaosa and Y.~Tokura, Nat. Nanotech. {\bfseries 8},  899 (2013).

\bibitem{zhang2014hidden}
X.~Zhang, Q.~Liu, J.-W. Luo, A.~J. Freeman, and A.~Zunger, Nat. Phys.
  {\bfseries 10},  387 (2014).

\bibitem{Yanase:JPSJ.83.014703}
Y.~Yanase, J. Phys. Soc. Jpn. {\bfseries 83},  014703 (2014).

\bibitem{Hayami_PhysRevB.90.024432}
S.~Hayami, H.~Kusunose, and Y.~Motome, Phys. Rev. B {\bfseries 90},  024432
  (2014).

\bibitem{Hayami_PhysRevB.90.081115}
S.~Hayami, H.~Kusunose, and Y.~Motome, Phys. Rev. B {\bfseries 90},  081115
  (2014).

\bibitem{Hayami_doi:10.7566/JPSJ.84.064717}
S.~Hayami, H.~Kusunose, and Y.~Motome, J. Phys. Soc. Jpn. {\bfseries 84},
  064717 (2015).

\bibitem{Fu_PhysRevLett.115.026401}
L.~Fu, Phys. Rev. Lett. {\bfseries 115},  026401 (2015).

\bibitem{volkov1981macroscopic}
B.~Volkov, A.~Gorbatsevich, Y.~V. Kopaev, and V.~Tugushev, Zh. Eksp. Teor. Fiz
  {\bfseries 81},  742 (1981).

\bibitem{dzyaloshinsky1958thermodynamic}
I.~Dzyaloshinsky, J. Phys. Chem. Solids {\bfseries 4},  241 (1958).

\bibitem{moriya1960anisotropic}
T.~Moriya, Phys. Rev. {\bfseries 120},  91 (1960).

\bibitem{Starykh_PhysRevB.82.014421}
O.~A. Starykh, H.~Katsura, and L.~Balents, Phys. Rev. B {\bfseries 82},  014421
  (2010).

\bibitem{kaplan1983single}
T.~Kaplan, Zeitschrift f{\"u}r Physik B Condensed Matter {\bfseries 49},  313
  (1983).

\bibitem{Shekhtman_PhysRevB.47.174}
L.~Shekhtman, A.~Aharony, and O.~Entin-Wohlman, Phys. Rev. B {\bfseries 47},
  174 (1993).

\bibitem{sugita2015antisymmetric}
Y.~Sugita, S.~Hayami, and Y.~Motome, Physics Procedia {\bfseries 75},  419
  (2015).

\bibitem{comment_Sugita}
Y. Sugita, S. Hayami, and Y. Motome, unpublished.

\bibitem{Luttinger_PhysRev.70.954}
J.~M. Luttinger and L.~Tisza, Phys. Rev. {\bfseries 70},  954 (1946).

\bibitem{Spaldin:0953-8984-20-43-434203}
N.~A. Spaldin, M.~Fiebig, and M.~Mostovoy, J. Phys.: Condens. Matter {\bfseries
  20},  434203 (2008).

\bibitem{comment_exchange_int}
Note that other effects, such as a single-ion anisotropy and anistropic
  exchange interactions, should be taken into account in order to realize the
  FM ordering.

\bibitem{Melcher_PhysRevLett.30.349}
R.~L. Melcher, Phys. Rev. Lett. {\bfseries 30},  349 (1973).

\bibitem{MiyaharaJPSJ.81.023712}
S.~Miyahara and N.~Furukawa, J. Phys. Soc. Jpn. {\bfseries 81},  023712 (2012).

\bibitem{Miyahara_PhysRevB.89.195145}
S.~Miyahara and N.~Furukawa, Phys. Rev. B {\bfseries 89},  195145 (2014).

\bibitem{Iguchi_PhysRevB.92.184419}
Y.~Iguchi, S.~Uemura, K.~Ueno, and Y.~Onose, Phys. Rev. B {\bfseries 92},
  184419 (2015).

\bibitem{seki2015nonreciprocal}
S.~Seki, Y.~Okamura, K.~Kondou, K.~Shibata, M.~Kubota, R.~Takagi, F.~Kagawa,
  M.~Kawasaki, G.~Tatara, Y.~Otani, and Y.~Tokura, arXiv:1505.02868 ,  (2015).

\bibitem{xiao2012coupled}
D.~Xiao, G.-B. Liu, W.~Feng, X.~Xu, and W.~Yao, Phys. Rev. Lett. {\bfseries
  108},  196802 (2012).

\bibitem{Gitgeatpong_PhysRevB.92.024423}
G.~Gitgeatpong, Y.~Zhao, M.~Avdeev, R.~O. Piltz, T.~J. Sato, and K.~Matan,
  Phys. Rev. B {\bfseries 92},  024423 (2015).

\bibitem{Ressouche_PhysRevB.82.100408}
E.~Ressouche, M.~Loire, V.~Simonet, R.~Ballou, A.~Stunault, and A.~Wildes,
  Phys. Rev. B {\bfseries 82},  100408 (2010).

\bibitem{li2013coupling}
X.~Li, T.~Cao, Q.~Niu, J.~Shi, and J.~Feng, Proc. Natl. Acad. Sci. U.S.A.
  {\bfseries 110},  3738 (2013).

\bibitem{Sivadas_PhysRevB.91.235425}
N.~Sivadas, M.~W. Daniels, R.~H. Swendsen, S.~Okamoto, and D.~Xiao, Phys. Rev.
  B {\bfseries 91},  235425 (2015).

\bibitem{tomita2015high}
T.~Tomita, K.~Kuga, Y.~Uwatoko, and S.~Nakatsuji, J. Phys.: Conf. Ser.
  {\bfseries 592},  012019 (2015).

\end{thebibliography}

\end{document}